\documentstyle[12pt,fleqn,aaspp4]{article}
\newcommand{\be}{\begin{equation}}
\newcommand{\ee}{\end{equation}}

\begin{document}

\title{Empirical Calibration of {\it Hubble Space Telescope} WFPC2 F606W and F814W Photometry}

\author{Samir Salim}
\affil{E-mail: samir@astronomy.ohio-state.edu}
\author{Andrew Gould\altaffilmark{1}\altaffiltext{1}{Alfred P.\ Sloan Foundation 
Fellow}} 
\affil{E-mail: gould@astronomy.ohio-state.edu}
\affil{Ohio State University, Department of Astronomy, 
174 West 18th Ave., Columbus, OH 43210}

\begin{abstract}

	Using the ground-based $V$ and $I$ photometry of a sample of stars from the Groth Strip, we obtain the first empirical calibration of the F606W and F814W {\it Hubble Space Telescope} WFPC2 filters for $0.5<V-I<4.5$. We present results in the form of corrections that need to be applied to the two {\it synthetic} calibrations currently in use. Both calibrations are found to require corrections to zero points and color-terms.

\end{abstract}
\clearpage

\section{Introduction}

	The {\it Hubble Space Telescope (HST)} Wide Field Planetary Camera 2 (WFPC2) has been a key astronomical instrument for the last five years and has produced a tremendous quantity of scientific data. Nevertheless, there have been few attempts to calibrate its set of filters in terms of standard Johnson/Cousins photometry. In particular, the currently used calibrations of WFPC2 F606W and F814W filters (roughly corresponding to {\it V} and {\it I} standard filters) rely entirely on {\it synthetic} transformations. 

	The most widely used calibration is that by Holtzman et al.\ (1995 H95). They give a synthetic calibration for F606W based on convolutions of WFPC2 response curves with stellar spectra from the Bruzual, Persson, Gunn \& Stryker atlas \texttt{(http://www.stsci.edu/ftp/cdbs/cdbs2/grid/bpgs/)}. While H95 do give an empirical calibration for F814W, it does not match the synthetic calibration, indicating that either the {\it I} or the F814W passband used by H95 may be incorrect. Also, since the stars observed by H95 (Landolt 1983, 1992a, 1992b standards) were ``not very red'', the authors suggest that for red stars one should switch from the empirical transformations to the synthetic ones. H95 give first and second order color-terms for all transformations. The synthetic transformations are given for two ranges of $V-I$ color.

	Bahcall et al.\ (1994 BFGK) give another synthetic calibration. They transformed F606W and F814W magnitudes into {\it V} and {\it I} using theoretical throughput and response curves and convolving them with Gunn \& Stryker (1983) standard spectra. They present piecewise linear fits for three ranges of $V-I$ colors. This fit extends to a color of $V-I=5$.

	A comparison between H95 and BFGK transformations indicates significant disagreement. In $V$, there is an overall offset of $\sim 0.12$ mag, with a slight color dependence. In $I$, the transformations give similar results for $V-I<3$, but start to differ for $V-I>3$, with the difference becoming 0.08 mag at $V-I=4$. This discrepancy strengthens the need for an empirical ground-based calibration, which one would want in any case. Any such calibration should include red stars ($V-I>3$). This is particularly important since star counts of faint field stars include many M dwarfs (Gould et al.\ 1997 and references therein). Moreover, all the synthetic calibrations for the red stars were based on spectra of M giants and not M dwarfs. Any uncertainty in the calibration of these red stars will propagate to the derived mass function of M dwarfs, leading to a less accurate mass census of the Galaxy.

	Here we present a ground-based calibration based on stars observed in a portion of the Groth Strip, 28 contiguous WFPC2 fields. We find deviations from both the H95 and the BFGK transformations. We present best fits to the residuals between each of these two transformations and the observed photometry, and we parameterize the corrections that should be applied to H95 and BFGK relations.

\section{Observations and Data Reduction}

	The observations were centered on one part of the Groth Strip (J2000.0 $\alpha= 14{^h} 17{^m} 04{^s}, \delta= +52^{\circ} 21' 55''$), spanning approximately 9 original WFPC2 fields ($\sim 36'$ square). We chose the Groth Strip because, in contrast to most WFPC2 fields, the size of the field is well-matched to ground-based CCDs and because photometry of its stellar contents was already available (Gould et al.\ 1997). In order to get a good signal-to-noise ratio for the very faint red stars (M dwarfs), only one CCD field ($7.\hskip-2pt'56$ square) was observed. We selected the field so as to include as many red ($V-I>3$) stars as possible.

	The observations were performed on the Hiltner 2.4-m telescope at the MDM Observatory, using the Echelle CCD camera with SITE $2048\times 2048$ detector. Of the two nights alloted, one was lost due to weather and equipment problems. The other night was photometric and we obtained a total of 7200 s in $V$ and 6000 s in $I$. Photometry was performed to limiting magnitudes of $V\sim 25.1$ and $I\sim 23.6$. We observed 15 Landolt (1992a) standards (including 5 red standards) and took twilight flats and bias frames. 

	We reduced the images using the {\it ccdred} package in {\it IRAF}. The bias images were not used because they showed evidence of a light leak (they were taken in daytime). We therefore debiased the images using the overscan region. The twilight flats were scaled and combined to produce flat fields which were normalized by their modes. The solution for the standards was based on $4''$ radius apertures, as they provided the best fit. The extinction coefficients we used come as a part of the solution for the standards ($Q_V=0.18\pm0.01, Q_I=0.08\pm0.01$). To check these values we measured bright stars in the observed field at various airmasses and obtained ($Q_V=0.17\pm0.01, Q_I=0.11\pm0.01$). The two sets of coefficients are in reasonable agreement. (The difference in $Q_I$ can produce a systematic error of only 0.002 mag). We then aligned and combined subsets of images grouped  according to the measured seeing. We thereby obtain two images in $V$ and one in $I$. Bad pixels were rejected using a mask. The seeing for the two combined images in $V$ is $0.\hskip-2pt''9$ and $1.\hskip-2pt''3$, and in $I$ it is $0.\hskip-2pt''8$. The aperture radius for field star photometry was chosen to equal 1.0 full width at half maximum seeing diameter (thus corresponding to the above values) as it proved to give the best consistency when comparing measurements of faint stars in the two $V$ images. We performed all photometry using the {\it qphot} task in {\it IRAF}.

\section{Data Analysis and Results}

	Out of approximately 75 stars in our field that were measured on the WFPC2 images (8 of them red), we were able to detect some 50. This number was reduced to 34 stars in $V$ (4 of them red) and 40 stars in $I$ (all 8 red stars detected) after excluding stars that were contaminated by close-lying galaxies in the ground-based photometry (as cross-checked with {\it HST} images) and those that were saturated in the {\it HST} photometry.

\subsection{WFPC2 Chip-to-Chip Offsets}

	BFGK introduce chip-to-chip offsets that are to be applied to each of the chips 2, 3 and 4 of WFPC2 (designated as $\delta_n$, $n=2,3,4$). These offsets were determined from the individual response curves of these chips that the authors used to construct their transformations. In order to check the validity of these chip-to-chip offsets, we conducted two independent tests. First we noted that the WFPC2 fields in the Groth Strip are so  arranged that some stars are found on both chip 2 and chip 3. Therefore, we compared BFGK magnitudes of 27 stars that were located in these overlapping regions. There was a difference in BFGK magnitudes that closely matched the offset between chip 2 and 3 ($\delta_3 - \delta_2 = 0.12$ mag) that BFGK use in both $V$ and $I$ transformations. We thus show that the introduction of BFGK chip-to-chip offsets was not needed, and rule them out at the $3\sigma$ confidence level. If any chip-to-chip offsets are present, we see no evidence that they are greater than couple hundreds of a magnitude, which is consistent with the difference in zero point values for different chips as listed in SYNPHOT table of zero points \texttt{(http://www.stsci.edu/ftp/instrument\_news/WFPC2/Wfpc2\_phot/wfpc2\_photlam.html)}.

	Second, we measured the mean differences between the BFGK {\it HST} $V$ and $I$ magnitudes and our ground-based magnitudes separately for each chip. With one exception (see below), we found mean differences that are similar to the offsets adopted by BFGK. That is, our results again show that BFGK offsets are not needed. However, for the $I$ filter on chip 4 this pattern failed to hold, and is off by 0.15 mag compared to $V$. If real, this effect would seem to suggest that different chips have substantially different spectral response. We consider this unlikely. See \S\ 3.2 for further discussion.

	In brief, we conclude that no chip-to-chip offsets of the BFGK type should be used.

\subsection{Photometric Transformations}

	Here we present our best fits to the residuals between our ground-based data and the magnitudes from the H95 and BFGK transformations (Figs. 1-4). To obtain the fits, we weighted the data points by the photometric errors scaled up to produce reduced $\chi^2\sim 1$. The zero points and errors were obtained from fits centered at $V-I=2$, in order to reduce correlation between the zero point and slope (color-term) errors. The quality of the data are sufficient to construct only linear fits.

	The corrections to H95 calibration are,
\begin{equation}
V = V_H + \Delta V_H,\qquad  
\Delta V_H=-0.092(\pm 0.007)-0.055(\pm 0.012)[(V-I)-2],
\end{equation}
\begin{equation}
I = I_H + \Delta I_H,\qquad    
\Delta I_H=-0.005(\pm 0.009)-0.028(\pm 0.015)[(V-I)-2],
\end{equation}
where $V_H$ and $I_H$ are the transformations given by equation (9) in H95, with zero points and first and second order color-terms given in Table 10 of H95.

	The corrections to the BFGK calibration are,
\begin{equation}
V = V_B + \Delta V_B,\qquad 
\Delta V_B=0.036(\pm 0.006)-0.051(\pm 0.010)[(V-I)-2],
\end{equation}
\begin{equation}
I = I_B + \Delta I_B,\qquad 
\Delta I_B=-0.017(\pm 0.011)-0.015(\pm 0.018)[(V-I)-2],
\end{equation}
where $V_B$ and $I_B$ are transformations given by equation (2.1) in BFGK, but with $\delta_n=0$, as previously explained. The BFGK zero points and color-terms are given in the text of BFGK.

	First, we notice that there are zero point offsets (corrections) compared to both H95 and BFGK photometries. It is especially significant in $V$ (for H95 it is $-0.09$ mag and for BFGK 0.04 mag at $V-I=2$).The zero point corrections in $I$ ($-0.01$ mag for H95 and $-0.02$ mag for BFGK, also at $V-I=2$) are much smaller. The H95 zero point in $I$ was verified recently by Garnavich et al.\ (1998) who report $I-I_H=0.00\pm0.04$.

	Of equal importance, there are corrections to the color-terms, which are especially significant in $V$. We find similar $V$ corrections for the H95 and the BFGK transformations. This is expected since the two calibrations agree to within $\pm0.04$ mag after subtracting the zero point offsets. In $I$, by contrast, H95 and the BFGK do not agree on the slope, so the slope corrections differ. However, the corrections of the color-term in $I$ are not as statistically significant as they are in $V$.

	Due to the discrepancy of chip-to-chip offset of chip 4 in $I$, we performed fits in which we excluded all the data points that came from WFPC2 chip 4. The fits in $V$ remain essentially unchanged, but the zero point offset in $I$ is increased by $\sim 0.07$ mag and the color-term correction is changed slightly for both H95 and BFGK. At this point we are not able to explain this effect or to trace its origin.

\section{Conclusion}

	We have compared our ground-based photometry of a sample of stars going to very red colors, to two synthetic photometric systems currently in use for {\it HST} F606W and F814W data. We find differences, especially significant in $V$, in both the zero point offsets, and the  color-terms. The color-term corrections are especially important if one wants to perform WFPC2 photometry on very red stars, as for example in the case of M dwarf star counts.

	We mention some of the consequences our corrections will have on the M dwarf star counts that Gould et al.\ (1997) performed using the BFGK transformations. Our corrections result in stars actually having a slightly redder $V-I$ colors. However, the trend of this color correction is such that it becomes zero at $V-I=3.5$, and is blue afterwards. Correction of color will shift the features of the M dwarf luminosity function (LF) ($M_V$ absolute magnitudes are determined from the $V-I$ color). The peak of the LF will be shifted by about 0.1 mag towards the fainter magnitudes. The change in apparent and absolute magnitudes will shorten the scale heights of the stars at the peak of the LF by $\sim 3\%$. The red end cut-off used to select the sample of M dwarfs will not be significantly changed. 

	Our photometry is not precise enough to determine whether H95 or BFGK is better to use as a starting point. In any event, there are significant corrections that must be applied to both. However, the BFGK system is simpler in that it is piecewise linear, and there does not appear to be sufficient information at the present time to justify the H95 piecewise quadratic transformations.

{\bf Acknowledgements}:  
We thank C. Flynn for assisting us with WFPC2 Groth Strip images. We thank J. N. Bahcall and C. Flynn for helpful comments on the manuscript. This work was supported in part by grant AST 97-27520 from the NSF.

\clearpage

\begin{figure}
\caption[junk]{\label{fig:one}
Fit to the residuals between our ground-based and H95 $V$ magnitudes. Data points with the errors $>0.25$ mag are plotted as open circles without error bars to avoid clutter.
}
\end{figure}

\begin{figure}
\caption[junk]{\label{fig:two}
Fit to the residuals between our ground-based and H95 $I$ magnitudes. Data points with the errors $>0.60$ mag are plotted as open circles without error bars to avoid clutter.
}
\end{figure}

\begin{figure}
\caption[junk]{\label{fig:three}
Fit to the residuals between our ground-based and BFGK $V$ magnitudes. Data points with the errors $>0.25$ mag are plotted as open circles without error bars to avoid clutter.
}
\end{figure}

\begin{figure}
\caption[junk]{\label{fig:four}
Fit to the residuals between our ground-based and BFGK $I$ magnitudes. Data points with the errors $>0.60$ mag are plotted as open circles without error bars to avoid clutter.
}
\end{figure}


\clearpage

\begin{references}
\reference{BFGK} Bahcall, J.\ N., Flynn, C., Gould, A. \& Kirhakos, S.\ 1994, \apj, 435, L51 (BFGK)
\reference{fl} Flynn, C., Gould, A. \& Bahcall J.\ N.\ 1996, \apj, 466, L55
\reference{Gar} Garnavich, P.\ M.\ et al.\ 1998, \apj, 493, L53
\reference{Gould} Gould, A., Bahcall, J.\ N. \& Flynn, C.\ 1997, \apj, 482, 913
\reference{GS} Gunn, J.\ E. \& Stryker, L.\ L.\ 1983, \apjs, 52, 121
\reference{H95} Holtzman, J.\ A.\ et al.\ 1995, PASP, 107, 1065 (H95)
\reference{L1} Landolt, A.\ U.\ 1983, \aj, 88, 439
\reference{L2} Landolt, A.\ U.\ 1992a, \aj, 104, 340
\reference{L3} Landolt, A.\ U.\ 1992b, \aj, 104, 372
\end{references}
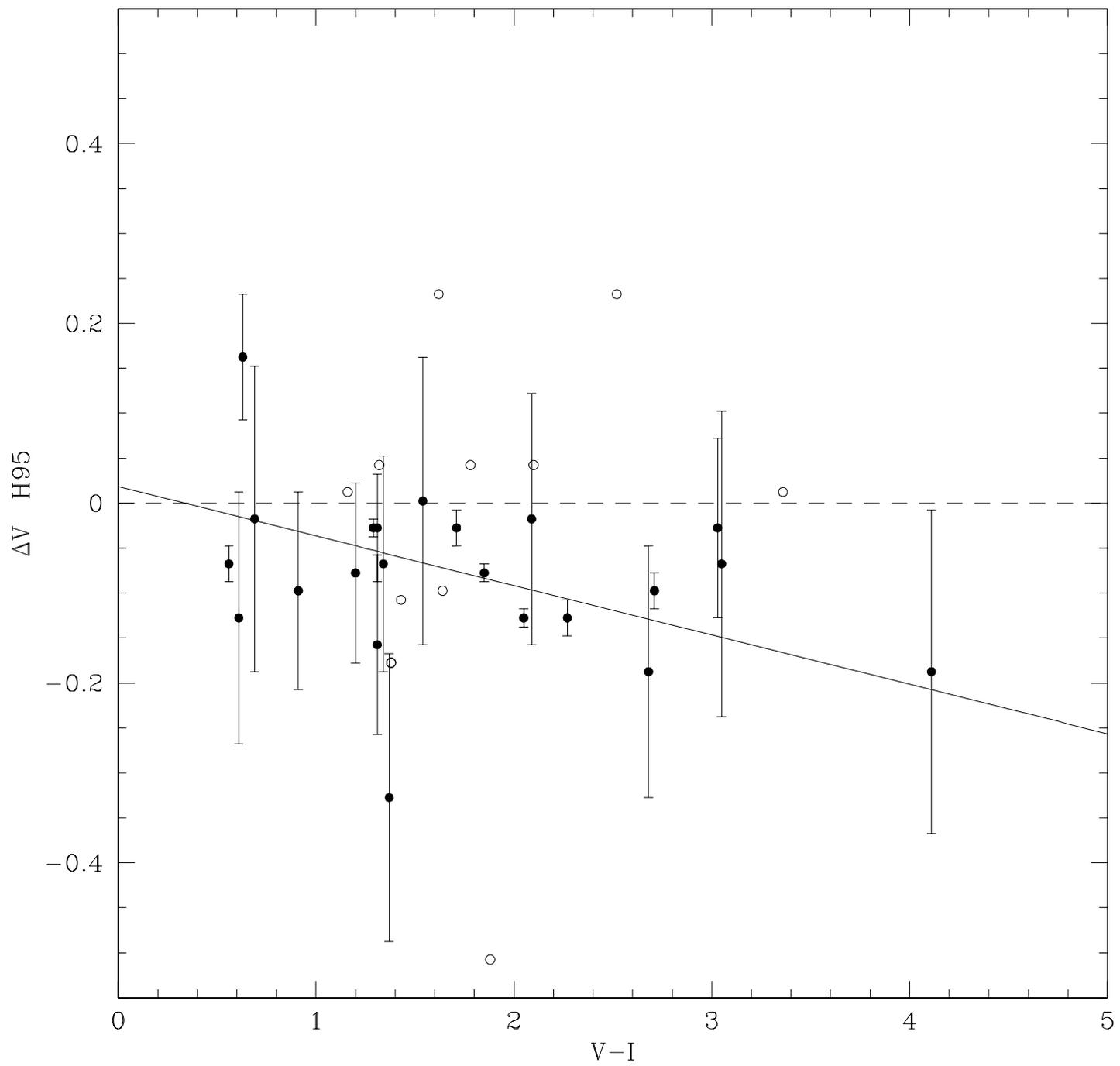
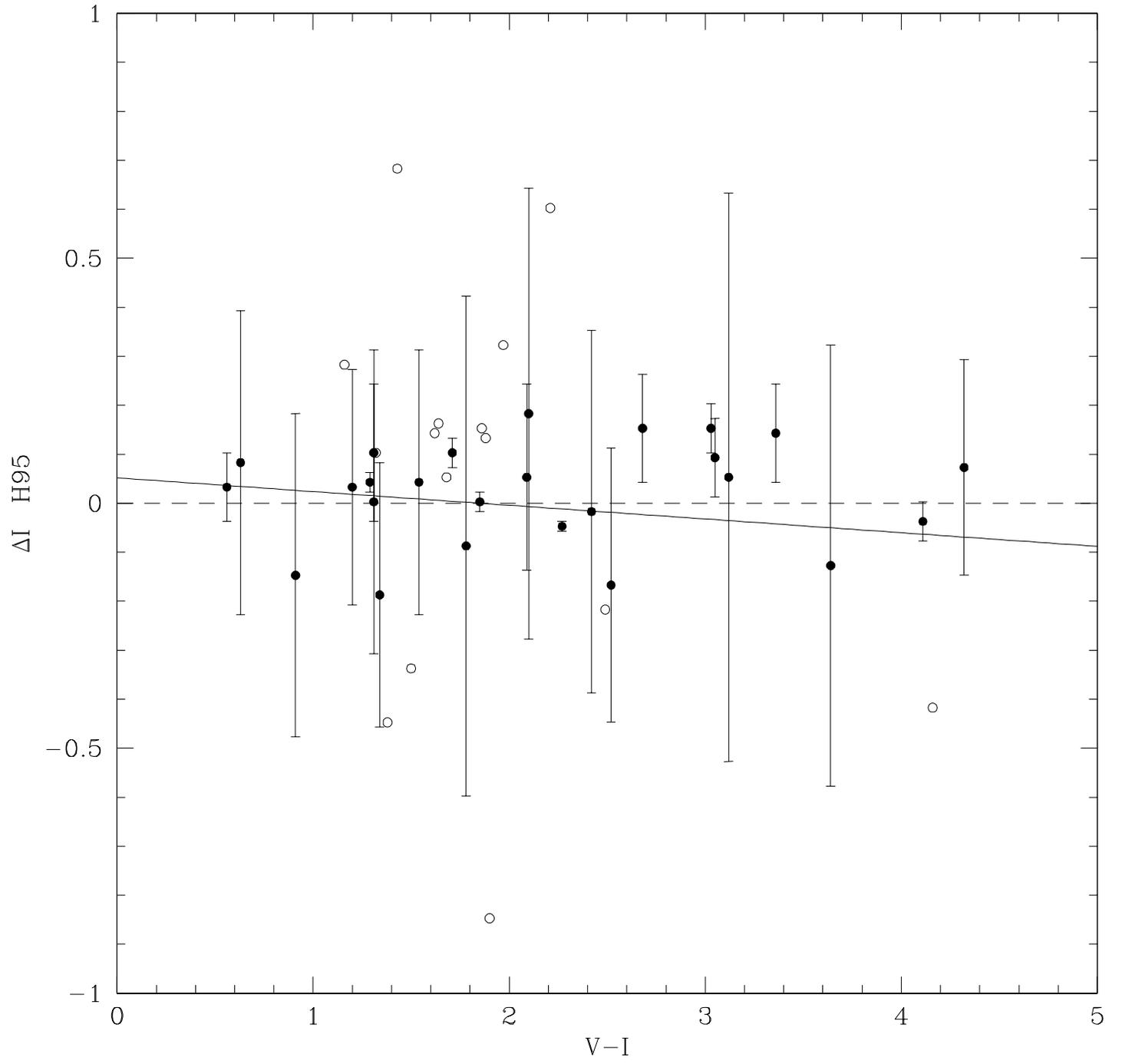
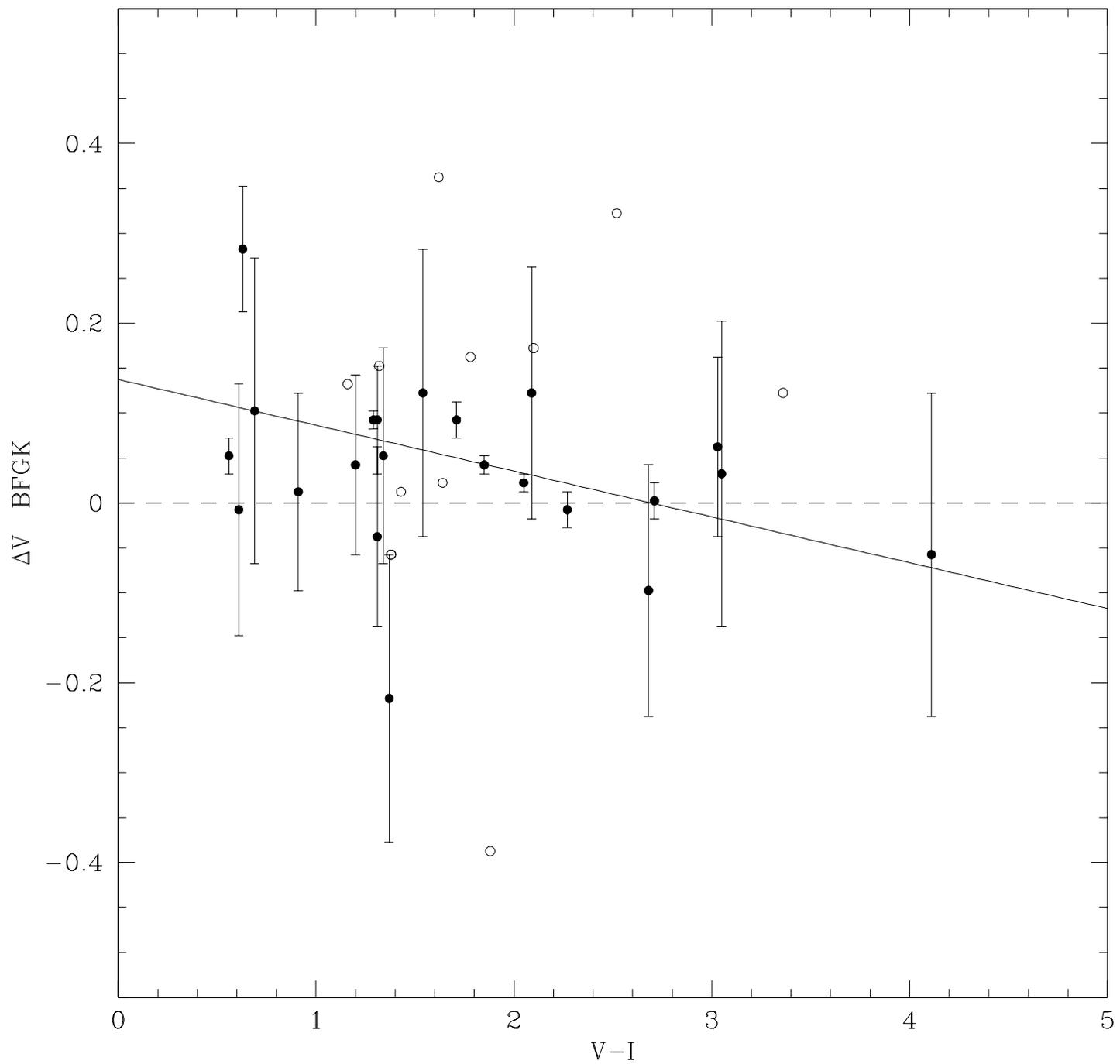
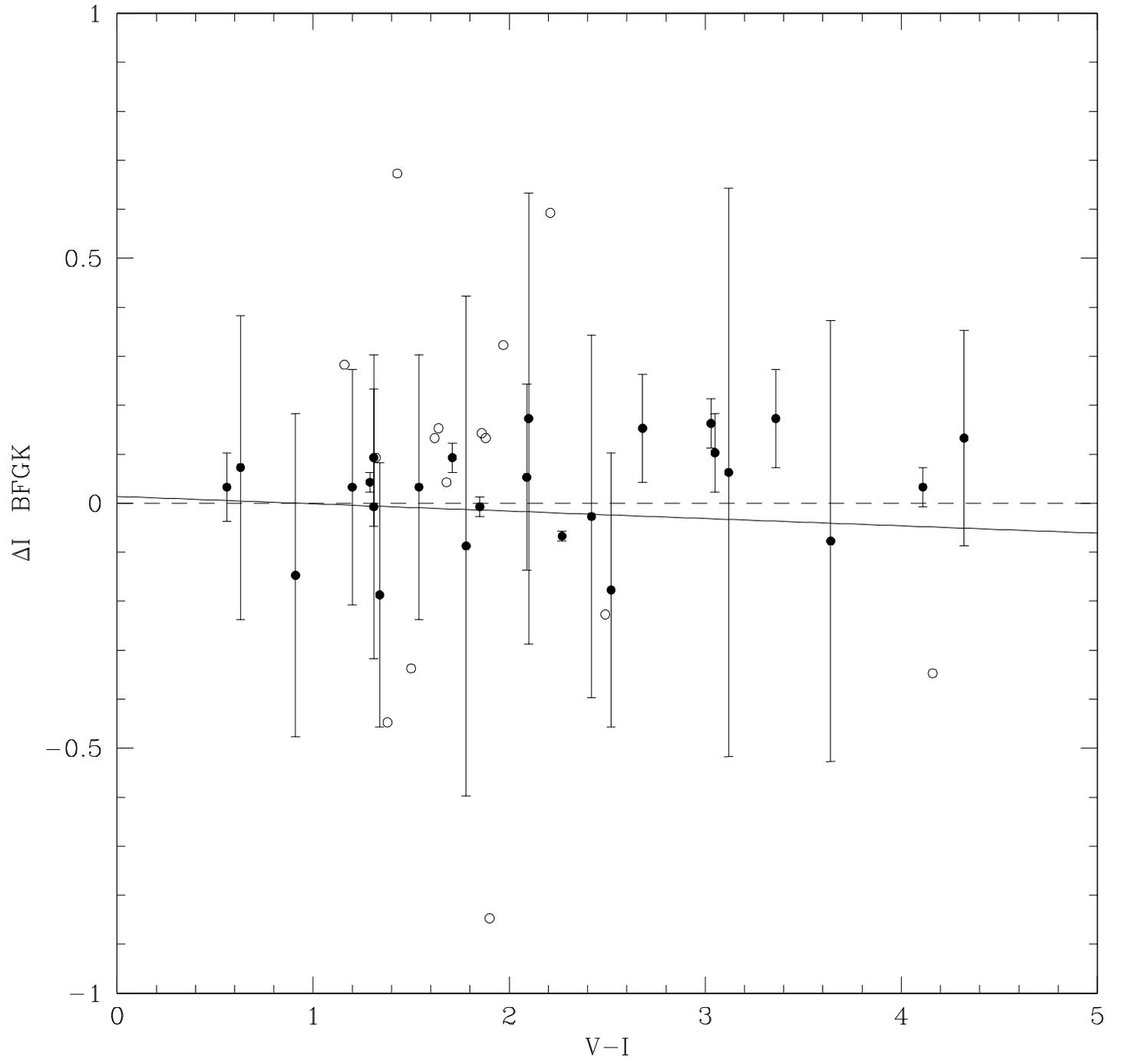
\end{document}